# Modeling Transport and Mechanics in Anion Exchange Membranes to Optimize Continuous $CO_2$ Separations

Samaneh Bayati[1], Jennifer Wade[1*], Sara Sarbaz[1], Heidi Feigenbaum[1], Horacio Lopez-Marques[2], Winston Wang[2]

(1) Northern Arizona University, Flagstaff, Arizona
(2) University of Texas at Austin, Austin, Texas

**Abstract**

Water transport in charged polymer membranes is strongly influenced by hydration-dependent transport properties and deformation of the polymer network. Under water-activity gradients, spatial variations in water concentration induce swelling and mechanical stresses, while the resulting deformation and mechanical state can, in turn, influence water transport. Capturing this bidirectional interaction is therefore important for predicting membrane behavior under nonuniform hydration conditions.

This study develops a transient finite element model of water transport and mechanical deformation in a Fumasep FAA-3-50 anion exchange membrane. The framework incorporates concentration-dependent water diffusivity, hygroscopic swelling, and the nonlinear viscoelastic response of the polymer matrix. Uncoupled, one-way coupled, and fully coupled formulations are compared to distinguish the effects of hydration-dependent diffusivity from those arising specifically from mechanical coupling. In the fully coupled formulation, local water concentration governs swelling and mechanical response, while the resulting deformation, evolving membrane configuration, and mechanical pressure feed back into water transport.

Simulations reveal spatially and temporally nonuniform swelling and deformation that follow the evolving water-concentration field. Comparison with the uncoupled concentration-dependent diffusion model shows that mechanical feedback modifies the predicted water-transport response, although transport remains predominantly diffusion-driven. The fully coupled model is further validated against steady-state water-transport measurements under variable water-activity gradients and captures the experimentally observed hydration dependence of water transport, with the closest agreement at intermediate hydration conditions. These results demonstrate the importance of resolving the bidirectional interaction among hydration, swelling, deformation, and transport when modeling hydrated anion exchange membranes and provide a framework for evaluating membrane behavior in moisture-facilitated transport processes.

## 1. Introduction

Charged polymer membranes have garnered significant attention for applications including water electrolysis, fuel cells, and gas separation. Their performance is strongly influenced by chemo-mechanical behavior, particularly the interaction between chemical transport and mechanical deformation. Water transport is especially important in ion-containing polymer membranes because spatial gradients in water activity can develop during operation and affect molecular transport, ionic conductivity, and the mechanical state of the polymer. In proton-conducting fuel-cell membranes, for example, through-plane water-concentration profiles arise from water-activity and electrochemical gradients, making water distribution an important determinant of membrane transport and device performance [1]. Water activity is also important in membrane separations because changes in hydration can alter species mobility and the local chemical environment through which transport occurs.

In ion-containing membranes, water-activity gradients can play an even more direct role by coupling water transport with ionic and reactive transport mechanisms. This behavior is particularly relevant to dilute $CO_2$ separations. Wade et al. [2] demonstrated that a sustained water-vapor gradient across a commercial anion exchange membrane (AEM) can drive continuous $CO_2$ transport through moisture-sensitive bicarbonate chemistry, establishing a moisture-driven mechanism for membrane-based $CO_2$ pumping. Related work has demonstrated continuous $CO_2$ capture from dilute sources using a migration-assisted moisture-gradient process that couples humidity, ionic transport, and carbonate chemistry [3]. More recent studies of moisture-swing anion-exchange materials have further shown that polymer chemistry, ionic functionality, water uptake, and material structure influence humidity-dependent $CO_2$ capture behavior [4,5]. Together, these studies establish hydration and water activity as important variables governing transport and separation behavior in ion-containing polymers.

Hydration also alters the mechanical state of these materials. For Fumasep FAA-3 specifically, Sarbaz et al. [6] experimentally demonstrated humidity-dependent swelling and changes in stiffness, strength, plastic deformation, and stress-relaxation behavior and developed a nonlinear viscoelastic constitutive model to describe the membrane response under varying environmental conditions. Thus, water uptake affects not only transport within the membrane but also deformation of the polymer matrix. The interaction can be bidirectional: local water concentration governs hygroscopic swelling and the mechanical state of the membrane, while the resulting deformation and stress state can, in turn, modify the environment through which water is transported. Understanding this coupling is therefore important for predicting membrane behavior under nonuniform hydration conditions.

The coupling between chemical transport and mechanical deformation in soft materials has been extensively studied, with foundational theories providing the basis for modern computational approaches. Narayan and Anand [7] developed a coupled electro-chemo-mechanical theory for polyelectrolyte gels, focusing on their swelling response to chemical stimuli. Their framework described the interplay among ionic transport, electrostatics, and mechanical stresses and was numerically implemented using the finite element method. Simulations demonstrated the transient swelling response of anionic and cationic polyelectrolyte gels under variations in environmental pH and ionic strength. Similarly, Hong et al. [8] proposed a theory for coupled diffusion and large deformation in polymeric gels, emphasizing solvent migration as a driver of mechanical deformation. They distinguished rapid local molecular rearrangement from the slower long-range migration of solvent molecules through the polymer network, with the latter enabling changes in both shape and volume and being governed by solvent diffusion.

Building on these foundations, finite element methods (FEM) have been widely applied to coupled transport and deformation in hydrated polymeric materials. Bouklas et al. [9] developed a nonlinear, transient FEM framework for coupled solvent diffusion and large deformation in hydrogels, capturing both transient and equilibrium behavior. Ballhause and Wallmersperger [10] developed chemo-electro-mechanical FEM simulations of polyelectrolyte gels subjected to changes in the salt concentration of the surrounding solution. Their model resolved the coupled evolution of concentration, electric potential, and displacement and related the resulting deformation to chemical and osmotic interactions between the gel and its environment. Shi and Wallmersperger [11] subsequently developed a finite-strain chemo-poro-mechanical framework for chemically stimulated fluid permeation in polymer gels. Their COMSOL implementation incorporated coupled chemical, electrical, porous-medium, and mechanical effects and was demonstrated for the swelling of an acrylamide gel under chemical stimulation.

Coupled formulations have also been developed for electroactive ionic polymers and hydrated polymer membranes. Bosnjak et al. [12] used FEM to investigate coupled electrochemical and mechanical behavior in soft ionic materials and ionotronic devices, including deformation associated with electrically driven ion redistribution. Rossi et al. [13] introduced a three-dimensional continuum electro-chemo-mechanical model for hydrated polymer membranes incorporating mechanics, water uptake, ion transport, and electrostatics. In their formulation, the deformation was separated into an elastic contribution and an electrochemically induced distortion associated with changes in the ionic state of the material. This concept is relevant to hydrated ion-containing membranes because changes in local composition can modify the preferred configuration of the polymer network, directly linking species redistribution to deformation rather than treating mechanics as independent of transport.

Kink et al. [14] modeled the mechanical behavior of proton exchange membranes in water electrolyzers using FEM to predict stresses and strains under mechanically and environmentally relevant conditions. Their simulations employed a Nafion® 117 membrane and an elastic-viscoplastic constitutive model. Similarly, Chester and Anand [15] developed a continuum-mechanical theory describing fluid permeation and large deformation, including swelling and squeezing, in elastomeric gels. Their work addressed three-dimensional swelling equilibrium and transient problems involving free swelling, consolidation of pre-swollen gels, and pressure-driven solvent diffusion through elastomeric membranes. Narayan et al. [16] subsequently introduced a thermodynamically consistent large-deformation electro-chemo-elasticity theory for ionic polymer–metal composites. The framework coupled the evolution of electric potential, hydrated-cation concentration, and deformation of the polymer matrix and was demonstrated for electroactive soft-actuator applications. Collectively, these studies demonstrate the importance of coupled transport–mechanical frameworks for describing soft ion-containing polymers subjected to chemical, hydration, and electrical stimuli.

While these studies establish important foundations for chemo-mechanical coupling in polyelectrolyte gels, hydrogels, and related soft ionic materials, their direct extension to AEMs requires consideration of differences in material structure and operating conditions. Polyelectrolyte gels and hydrogels generally consist of crosslinked polymer networks containing a substantial solvent phase and can undergo large solvent-induced changes in volume. AEMs, in contrast, are dense ion-containing polymer membranes containing fixed cationic functional groups and mobile counterions, with water absorbed within the polymer matrix and associated aqueous domains. Consequently, changes in hydration can simultaneously alter ionic environments, water mobility, polymer-network dimensions, and mechanical properties. These distinctions motivate direct characterization and modeling of transport–mechanical coupling in AEMs rather than assuming that behavior established for highly swollen gels can be transferred directly to dense membrane systems.

For FAA-3, previous studies have separately established important hydration-dependent transport and mechanical behavior. Water mobility in AEMs exhibits a strong dependence on hydration, with changes in aqueous-domain dimensions influencing water diffusion [19], while the water uptake and mechanical response of FAA-3 vary substantially with environmental humidity [6]. In addition, moisture-driven $CO_2$ pumping through FAA-3 relies directly on maintaining a water-vapor gradient across the membrane [2]. Despite these advances, the bidirectional interaction between transient water transport and deformation in FAA-3 has not been explicitly resolved. In particular, it remains necessary to determine how spatially varying water concentration produces local swelling and mechanical fields and, conversely, how the resulting deformation and mechanical state affect water transport.

This work addresses this gap by developing a fully coupled, two-dimensional transient finite element model for water transport and mechanical deformation in Fumasep FAA-3-50. The framework explicitly captures both directions of coupling: the evolving local water concentration determines hygroscopic swelling and the mechanical response of the membrane, while swelling-induced deformation, changes in membrane geometry, and mechanical pressure feed back into the water-transport problem. This formulation is compared with an uncoupled concentration-dependent diffusion model using the same hydration-dependent water diffusivity, allowing the influence of mechanical coupling to be distinguished from the intrinsic concentration dependence of diffusivity. The model is further experimentally validated against steady-state water-transport measurements obtained under variable water-activity gradients.

By quantifying the interaction among hydration, swelling, deformation, and water transport, this study extends established chemo-mechanical approaches to an AEM subjected to water-activity gradients. The resulting framework provides mechanistic insight into when concentration-dependent diffusion alone is sufficient and when mechanical feedback must also be considered in predicting membrane transport. Although the present study focuses specifically on water transport rather than explicitly modeling $CO_2$ transport, resolving the hydration–mechanics interaction provides a physical foundation for understanding moisture-facilitated membrane processes in which water-activity gradients govern the membrane environment and influence separation behavior.

## 2. Materials and methods

### 2.1 Physical Description of the Membrane System

The behavior of hydrated anion exchange membranes (AEMs) is governed by the interaction between water transport and deformation of the polymer network. When a water activity gradient is imposed across the membrane, water diffuses from the high-activity side toward the low-activity side, producing spatial and temporal variations in water concentration. Water uptake causes hygroscopic swelling of the polymer network, generating deformation and internal stresses. The resulting changes in membrane geometry, volume, and stress state can, in turn, influence subsequent water transport. A coupled transport-mechanical framework is therefore adopted to describe the transient response of the hydrated membrane.

In this study, the FAA-3-50 membrane is modeled as a deformable continuum in which water transport and mechanical deformation are linked through hygroscopic swelling. Water transport is described by transient diffusion, while the membrane response is represented using the nonlinear viscoelastic Parallel Network Model previously developed and calibrated for FAA-3-50 by Sarbaz et al. [6]. Hygroscopic swelling provides the coupling mechanism by converting changes in water concentration into volumetric expansion of the polymer network. The resulting deformation modifies the membrane geometry and internal stress state, thereby providing feedback to the diffusion process.

Three modeling approaches are considered to evaluate the influence of mechanical coupling. In the uncoupled transport model**,** water diffusion is solved independently without accounting for membrane deformation. In the one-way coupled model, the calculated water concentration determines hygroscopic swelling and deformation, while the resulting mechanical response does not affect water transport. In the fully coupled transport-mechanical model, the transport and mechanical equations are solved simultaneously, allowing deformation-induced changes in membrane geometry, volume, and internal stress to influence water transport. Comparing these three formulations provides insight into the role of mechanical coupling in water transport through the membrane.

To isolate the effect of mechanical coupling, the uncoupled and fully coupled models employ the same concentration-dependent water diffusivity. Consequently, any differences between these models arise solely from deformation-induced feedback rather than differences in the assumed transport properties.

### 2.2 Governing Equations for Water Transport

Water transport through the membrane is driven by a gradient in water activity established between the two membrane surfaces. A difference in external relative humidity produces a corresponding difference in the equilibrium water concentration within the membrane, as determined by the experimentally measured water sorption isotherm. Consequently, the imposed humidity gradient generates a water concentration gradient that serves as the driving force for diffusion across the membrane thickness.

The transient distribution of water within the membrane is governed by species conservation:

$$\frac{\partial c_w}{\partial t} + \nabla \cdot \mathbf{j}_{\boldsymbol{w}} = S_w \quad (1)$$

where $c_w$ is the water concentration (mol/m$^3$), $\mathbf{j}_{\boldsymbol{w}}$ is the water flux (mol/m$^2$ s), and $S_w$ is a source term representing internal generation or consumption of water. Since no chemical reactions or phase change occur within the membrane during transport, $S_w = 0$.

Water transport is assumed to occur in response to gradients in chemical potential. Accordingly, the water flux is first expressed in its general thermodynamic form as Eq. (2),

$$\mathbf{j}_w = -m_w c_w \nabla \mu_w \quad (2)$$

where $m_w$ is the water mobility and $\mu_w$ is the chemical potential.

Assuming ideal solution behavior, the chemical potential of water can be expressed as:

$$\mu_w = \mu_w^0 + RTln\frac{c_w}{c_{w.ref}} + p\Omega_w \quad (3)$$

where $\mu_w^0$ is the reference chemical potential, R is the universal gas constant, T is the absolute temperature, $c_w$ is the local water concentration, $c_{w.ref}$ is the reference concentration, p is the internal mechanical pressure, and $\Omega_w$ is the partial molar volume of water.

Equation (3) represents the general expression for the chemical potential of water. In the uncoupled transport model, the mechanical contribution is neglected, whereas it is retained in the fully coupled formulation presented in Section 2.5.

For the uncoupled transport model, the mechanical contribution to the chemical potential is neglected by setting ∇p=0. Substituting the resulting chemical potential into Eq. (2), together with the mobility relation given by Eq. (4), yields Fick's law (Eq. (5)).

$$m_w = \frac{D(c_w)}{RT} \quad (4)$$

$$\mathbf{j}_w = -D(c_w)\nabla c_w \quad (5)$$

where D($c_w$) is the concentration-dependent water diffusivity.

Experimental studies have consistently shown that water diffusivity in anion exchange membranes increases nonlinearly with membrane hydration. Although diffusivity was not measured directly in the present work, previous studies have quantified the relationship between water diffusivity and equilibrium water concentration using pulsed-field gradient nuclear magnetic resonance (PFG-NMR) and quasi-elastic neutron scattering (QENS). Because the QENS measurements provide reliable diffusivity data over the hydration range considered in this study, the published data of Melchior et al. [19] were used to determine the concentration-dependent diffusivity.

The equilibrium water concentration corresponding to each relative humidity was determined from the membrane water sorption isotherm reported by Sarbaz et al. [6]. Therefore, increasing relative humidity results in a higher equilibrium water concentration within the membrane, which in turn leads to increased water diffusivity. Figure 1 presents the third-order polynomial fit relating the logarithm of the diffusivity to the equilibrium water concentration. This fitted relationship was implemented throughout all transport simulations.

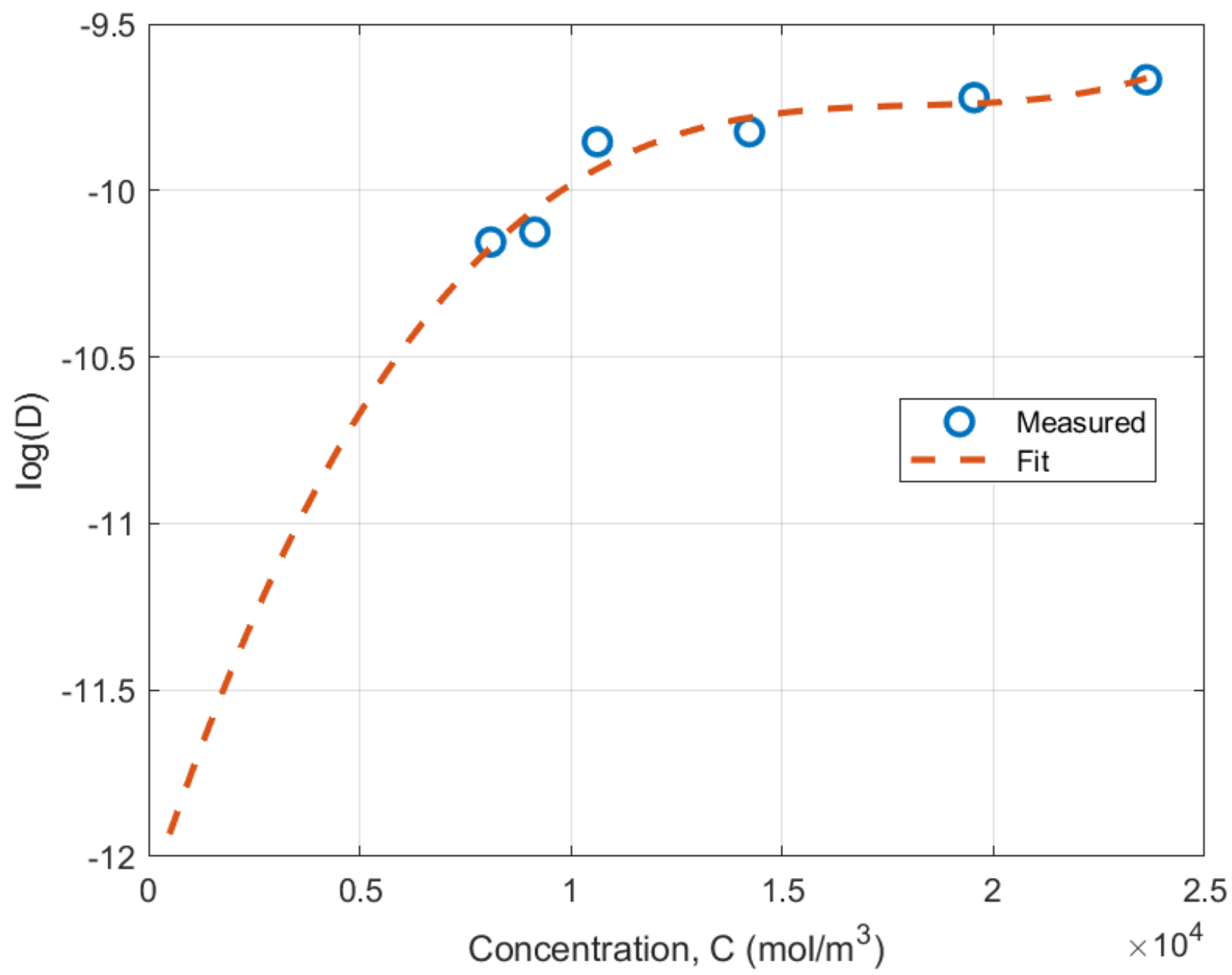

Figure 1. A third-degree polynomial fit of log D as a function of concentration based on QENS data [19].

The reported water diffusivity increases nonlinearly with membrane water content and decreases significantly at lower hydration levels. Melchior et al. attributed this reduction to the decreased thickness of the membrane's aqueous domains, which restricts water diffusion within the membrane [19].

Using a concentration-dependent diffusivity enables the model to capture the experimentally observed nonlinear transport behavior of hydrated membranes more accurately than assuming a constant diffusivity.

### 2.3 Mechanical Model

The mechanical response of the FAA-3-50 membrane is governed by the balance of linear momentum. In the current configuration, this balance is written as Eq. (6),

$$\rho \frac{\partial^2 \mathbf{u}}{\partial t^2} = \nabla \cdot \boldsymbol{\sigma} + \mathbf{f}_v \tag{6}$$

where $\rho$ is the membrane density, u is the displacement vector, $\boldsymbol{\sigma}$ is the Cauchy stress tensor, and $\mathbf{f}_v$ is the body force per unit volume. Equation (6) states that the acceleration of the membrane is produced by internal stress gradients and applied body forces.

The characteristic timescale of mechanical equilibration is much shorter than that of moisture transport. The deformation is therefore treated as quasi-static, and inertial effects are neglected. No body forces are applied in the present model, such that Eq. (6) reduces to Eq. (7),

$$\nabla \cdot \boldsymbol{\sigma} = 0 \tag{7}$$

The FAA-3-50 membrane is represented using the nonlinear, rate-dependent Parallel Network Model previously developed and calibrated for this material by Sarbaz et al. [6]. The model was selected because the experimental response of FAA-3-50 exhibits nonlinear elasticity, stress relaxation, and residual deformation that cannot be described using a purely elastic constitutive law. Reference [6] provides the experimental characterization and parameter-identification procedure used to establish the model.

As illustrated in Figure 2, the constitutive model consists of two parallel networks subjected to the same total deformation. Network A is a nonlinear elastic network represented by a large-deformation neo-Hookean model and describes the equilibrium stress response. Network B consists of a neo-Hookean elastic element in series with a Bergström–Boyce viscous flow element and captures the time-dependent response. The total Cauchy stress is obtained by adding the stress contributions from the two networks:

$$\boldsymbol{\sigma} = \boldsymbol{\sigma}_{\mathrm{A}} + \boldsymbol{\sigma}_{\mathrm{B}} \tag{8}$$

where $\boldsymbol{\sigma}_{\mathrm{A}}$ and $\boldsymbol{\sigma}_{\mathrm{B}}$ are the Cauchy stresses carried by Networks A and B, respectively.

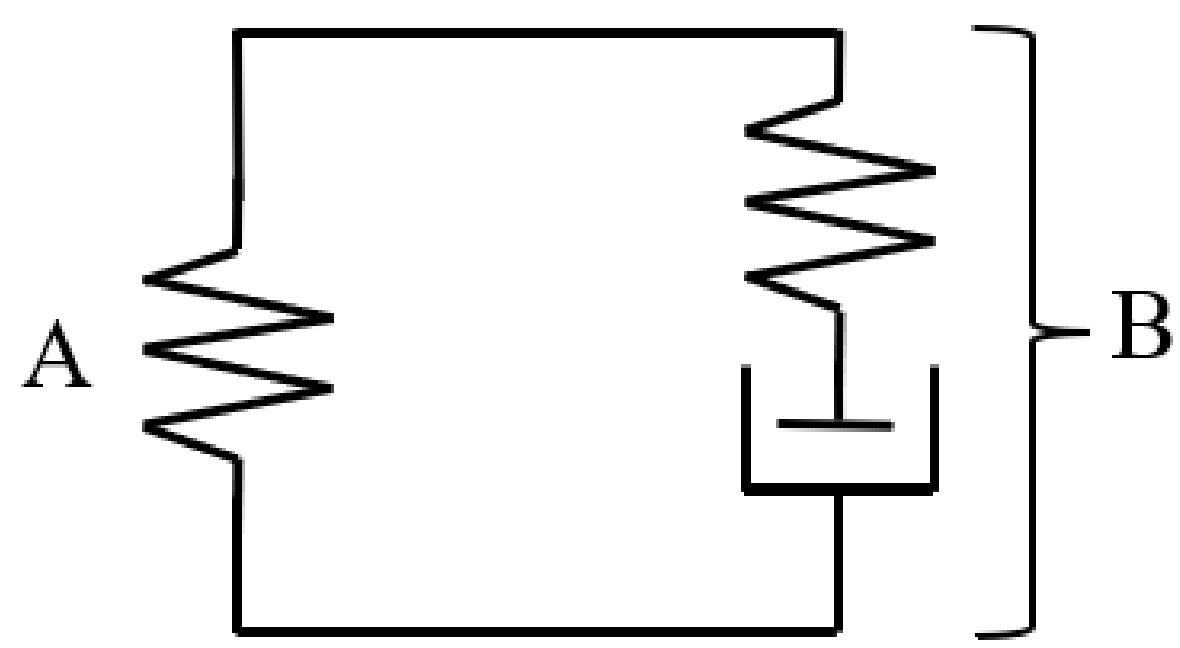

**Figure 2.** Schematic representation of the Parallel Network Model used to describe the nonlinear, time-dependent mechanical response of the FAA-3-50 membrane. Network A represents the equilibrium nonlinear elastic response, whereas Network B combines a nonlinear elastic element with a Bergström–Boyce viscous flow element to describe rate-dependent deformation.

Because the membrane undergoes hygroscopic expansion in addition to load-induced deformation, the total strain is decomposed into mechanical and hygroscopic contributions:

$$\varepsilon_{tot} = \varepsilon_{mech} + \varepsilon_{h} \tag{9}$$

where $\varepsilon_{tot}$ is the total strain, $\varepsilon_{mech}$ is the stress-producing mechanical strain, and $\varepsilon_h$ is the strain generated by water uptake. Thermal strain is omitted from Eq. (9) because each simulation is performed at a fixed temperature. Although the mechanical parameters retain the temperature dependence identified during the original material calibration, temperature does not vary within an individual simulation.

The deformation gradient is defined from the displacement field as

$$\mathbf{F} = \mathbf{I} + \nabla\mathbf{u} \tag{10}$$

where **F** is the total deformation gradient and **I** is the second-order identity tensor. The same total deformation is applied to both parallel networks. Within Network B, the deformation is additionally separated into elastic and viscous components according to the Bergström–Boyce formulation described in Ref. [6].

The mechanical properties of the two networks depend on the hydration state of the membrane. The shear moduli of Networks A and B are denoted by $G_1$ and $G_2$, respectively, and the effective flow resistance of the viscous element is denoted by $\hat{\tau}$. Following the calibration reported by Sarbaz et al. [6], these properties are prescribed as functions of the water concentration and temperature:

$$G_1(c_w.T) = -0 \cdot 00006\, c_w - 0 \cdot 1740\, T + 21 \cdot 7512 \tag{11}$$

$$G_2(c_w.T) = -0 \cdot 0162\, c_w - 0 \cdot 2802\, T + 306 \cdot 4701 \tag{12}$$

$$\hat{\tau}(c_w.T) = -0 \cdot 0048\, c_w + 0 \cdot 2355\, T + 80 \cdot 8866 \tag{13}$$

The concentration dependence in Eqs. (11) – (13) enables the model to capture hydration-induced changes in the mechanical behavior of the membrane. As the water concentration increases, the polymer network becomes more compliant, resulting in reduced stiffness and an increased tendency for deformation. The constitutive model developed by Sarbaz et al. [6] also incorporates temperature dependence through the calibrated material parameters. However, because all simulations in the present study are performed under isothermal conditions, the mechanical response varies only with the local water concentration.

The explicit calibrated expressions for $G_1$ , $G_2$, and $\hat{\tau}$, together with the constant Bergström–Boyce parameters, are summarized in Table 1. The constants ξ, $c_{BB}$ , and $m_{BB}$ control the strain and stress dependence of the viscous flow law. Subscripts are added to $c_{BB}$ and $m_{BB}$ to distinguish these constitutive constants from the water concentration used in the transport model.

**Table 1.** Material, transport, swelling, and geometric parameters used in the FAA-3-50 membrane simulations. Sources of the parameter values and hydration-dependent constitutive relationships are indicated.

| Category | Parameter | Description | Value or relationship | Units | Source |
|---|---|---|---|---|---|
| **Mechanical** | $\rho_{membrane}$ | Membrane density | 1300 | kg $m^{-3}$ | Wade et al. [2] |
| | E | Young's modulus | $6.25\times10^{8}$ | Pa | Sarbaz et al. [6] |
| | ν | Poisson's ratio | 0.30 | – | Sarbaz et al. [6] |
| | $G_1$ | Shear modulus of Network A | Eq. (11) | MPa | Sarbaz et al. [6] |
| | $G_2$ | Shear modulus of Network B | Eq. (12) | MPa | Sarbaz et al. [6] |
| | τ | Effective viscous flow resistance of Network B | Eq. (13) | MPa | Sarbaz et al. [6] |
| | ξ | Bergström–Boyce flow parameter | 0.078 | – | Sarbaz et al. [6] |
| | $c_{BB}$ | Stress-dependence parameter | −0.7 | – | Sarbaz et al. [6] |
| | $m_{BB}$ | Flow exponent | 11.28 | – | Sarbaz et al. [6] |
| **Swelling** | $\beta_h$ | Hygroscopic swelling coefficient | $5.13\times10^{-4}$ | $m^3$ $kg^{-1}$ | Sarbaz et al. [6] |
| | $M_w$ | Molar mass of water | 0.018015 | kg $mol^{-1}$ | Physical constant |
| | $c_{w,ref}$ | Reference water concentration | Equilibrium concentration at initial dry-side water activity | mol $m^{-3}$ | Sarbaz et al. [6] / sorption relationship |
| **Transport** | $D(c_w)$ | Concentration-dependent water diffusivity | Third-order polynomial fit, Fig. 1 | $m^2$ $s^{-1}$ | Melchior et al. [19] |
| | $\Omega_w$ | Molar volume of liquid water | $1.8\times10^{-5}$ | $m^3$ $mol^{-1}$ | Sarbaz et al. [6] |
| | R | Universal gas constant | 8.314 | J $mol^{-1}$ $K^{-1}$ | Physical constant |
| | T | Simulation temperature | 303.15 | K | Experimental condition |
| **Geometry** | L | Initial membrane thickness | 50 | µm | Model geometry |
| | W | In-plane computational length | 100 | µm | Model geometry |

The mechanical formulation described above determines how the membrane responds to a prescribed hygroscopic strain. The relationship between the local water concentration obtained from the transport model and the hygroscopic strain appearing in Eq. (9) is developed in Section 2.4.

### 2.4 Hygroscopic Swelling and One-Way Coupling

Water uptake produces expansion of the polymer network and provides the link between the transport and mechanical models. In the present formulation, moisture-induced expansion is modeled as an isotropic hygroscopic strain proportional to the local change in water concentration, as expressed in Eq. (14):

$$\varepsilon_h = \beta_h M_w \Delta c_w \boldsymbol{I} \tag{14}$$

where $\varepsilon_h$ is the hygroscopic strain tensor, $\beta_h$ is the hygroscopic swelling coefficient, $M_w$ is the molar mass of water, and $\boldsymbol{I}$ is the second-order identity tensor. The change in water concentration is

$$\Delta c_w = c_w - c_{w.ref} \tag{15}$$

where $c_w$ is the instantaneous local molar concentration of water obtained from the transport model and $c_{w.ref}$ is the reference water concentration.

The reference concentration $c_{w.ref}$ corresponds to the equilibrium water concentration at the initial hydration condition of the membrane and defines the strain-free reference state for hygroscopic swelling. In the present simulations, the membrane is initially equilibrated with the dry-side water activity; therefore, $c_{w.ref}$ is determined from the equilibrium concentration associated with that initial condition. Equation (14) gives zero hygroscopic strain when $c_w = c_{w.ref}$, positive swelling when the local water concentration exceeds the reference value, and contraction when the concentration falls below the reference value.

The hygroscopic swelling coefficient βh  was obtained from the experimental characterization and mechanical model calibration reported by Sarbaz et al. [6]. Its value and units are provided in Table 1 together with the remaining mechanical and transport parameters. Further details regarding its experimental determination are available in Ref. [6].

The hygroscopic strain is evaluated throughout the membrane using the transient local water concentration obtained from the transport model. Consequently, swelling-induced deformation evolves spatially and temporally as moisture diffuses through the membrane, producing nonuniform strain and stress distributions.

In the one-way coupled formulation, the transient water concentration field is transferred to the mechanical model through Eqs. (14) and (15). The resulting hygroscopic strain contributes to the total strain defined in Eq. (9), generating membrane deformation and internal stresses. However, the calculated mechanical response does not influence the transport equation. Consequently, the coupling is unidirectional, with water transport affecting the mechanical response while deformation has no feedback on water transport.

The one-way coupled model isolates the mechanical response induced by moisture uptake and provides a baseline for comparison with the fully coupled transport-mechanical formulation presented in Section 2.5.

### 2.5 Fully Coupled Transport-Mechanical Formulation

The fully coupled transport-mechanical formulation extends the one-way coupled model by allowing the evolving mechanical deformation to influence water transport. In addition to concentration gradients, deformation modifies the chemical potential through the development of internal mechanical stresses, thereby altering the thermodynamic driving force for diffusion. Consequently, the transport and mechanical governing equations are solved simultaneously, allowing the evolving deformation field to continuously influence the transport process throughout the transient simulation.

To account for the effect of hygroscopic swelling on the mechanical response, the total deformation gradient is multiplicatively decomposed into mechanical and swelling components according to

$$\mathbf{F} = \mathbf{F^m F^s} \tag{16}$$

where $\mathbf{F}$ is the total deformation gradient, $\mathbf{F^m}$ is the mechanical deformation gradient, and $\mathbf{F^s}$ is the swelling deformation gradient.

Assuming isotropic hygroscopic expansion, the swelling deformation gradient is expressed as

$$\mathbf{F^s} = \left(1 + \beta_h M_w\left(c_w - c_{w.ref}\right)\right)^{1/3} \mathbf{I} \tag{17}$$

where $\beta_h$ is the hygroscopic swelling coefficient introduced in Section 2.4, $M_w$ is the molar mass of water, $c_w$ is the local water concentration, $c_{w.ref}$ is the reference water concentration, and $\boldsymbol{I}$ is the identity tensor. The exponent $1/3$ distributes the volumetric hygroscopic expansion equally among the three principal directions under the assumption of isotropic swelling.

Unlike the uncoupled transport model presented in Section 2.2, where the mechanical contribution to the chemical potential is neglected, the fully coupled formulation retains the pressure term in Eq. (3). The internal mechanical pressure is obtained from the hydrostatic component of the stress field as

$$p = -\frac{1}{3} tr(\boldsymbol{F}^{m^T} \boldsymbol{T}_R \boldsymbol{F}^s) \tag{18}$$

where $\boldsymbol{T}_R$ is the first Piola–Kirchhoff stress tensor. The internal mechanical pressure represents the mean normal stress acting within the deforming polymer network. As the membrane swells, spatial variations in this pressure modify the local chemical potential of water and therefore influence the thermodynamic driving force for diffusion.

Substituting the complete chemical potential (Eq. (3)) into the general flux expression (Eq. (2)), together with the mobility relation given by Eq. (4), yields

$$\mathbf{j}_w = -D(c_w)\nabla c_w - \frac{D(c_w)\Omega_w}{RT} c_w \nabla p \tag{19}$$

where the first term represents classical Fickian diffusion driven by concentration gradients, whereas the second term accounts for changes in chemical potential arising from deformation-induced mechanical pressure gradients. Unlike bulk fluid convection, this second contribution represents a thermodynamic driving force associated with the evolving stress state of the polymer network.

Substituting Eq. (19) into the species conservation equation (Eq. (1)) gives the governing transport equation for the fully coupled formulation,

$$\frac{\partial c_w}{\partial t} + \nabla \cdot \left[-D(c_w)\nabla c_w - \frac{D(c_w)\Omega_w}{RT} c_w \nabla p\right] = 0 \tag{20}$$

which simultaneously accounts for concentration-driven diffusion and the influence of deformation-induced mechanical pressure on water transport.

To quantify the relative contribution of the mechanically induced transport term in Eq. (19), the pressure-dependent flux contribution is expressed in a drift-like form as

$$\mathbf{j}_{w.m} = c_w \mathbf{v}_m \tag{21}$$

where $\mathbf{v}_m$ is the effective mechanically induced drift velocity. Comparison with the pressure-gradient term in Eq. (19) gives

$$\mathbf{v}_m = -\frac{D(c_w)\Omega_w}{RT} \nabla p \tag{22}$$

Thus, $\mathbf{v}_m$ arises directly from the gradient in the internal mechanical pressure and does not represent bulk fluid convection. The relative magnitude of mechanically induced transport compared with diffusion was evaluated using a Peclet-type number,

$$Pe_m = \frac{|\mathbf{v}_m| L}{D(c_w)} \tag{23}$$

where L is the membrane thickness and D is the characteristic water diffusivity. For the present system, $L \approx 5 \times 10^{-5}\ m$ and $D \approx\ 10^{-12} - 10^{-10}\ {m^2}/{s}$. Substituting Eq. (22) into Eq. (23) gives

$$Pe_m = \frac{\Omega_w L}{RT} |\nabla p| \tag{24}$$

Using the characteristic mechanical-pressure gradient obtained from the simulations yields $Pe_m < 1$, indicating that the mechanically induced contribution is substantially smaller than the concentration-driven diffusive contribution. Moisture transport therefore remains diffusion-dominated even when swelling-induced stresses develop within the polymer bulk, and the direct contribution of the stress-assisted transport term to the overall water flux is negligible.

Nevertheless, transport-mechanical coupling remains important because swelling-induced deformation modifies the membrane geometry, local volume, and concentration distribution, thereby altering the effective diffusion pathways. These indirect effects influence both the transient evolution and steady-state water transport behavior even when the direct pressure-gradient contribution to the flux is small.

### 2.6 Numerical Implementation

The governing transport and mechanical formulations described in Sections 2.2–2.5 were implemented using the finite element method in COMSOL Multiphysics®. A two-dimensional transient model was developed to represent a cross-section of the FAA-3-50 membrane and to resolve the spatial and temporal evolution of water concentration and mechanical deformation under an imposed water-activity gradient. The computational domain had an initial membrane thickness of 50 μm and an in-plane length of 100 μm. The same geometry, material properties, and environmental conditions were used for the uncoupled, one-way coupled, and fully coupled simulations to enable direct comparison among the three formulations.
For the uncoupled and one-way coupled formulations, water transport was described using the concentration-dependent diffusion model introduced in Section 2.2. For the fully coupled formulation, the Stabilized Convection-Diffusion Equation (scdeq) interface was used to implement the modified governing transport equation derived in Section 2.5. This equation-based formulation allowed the mechanically induced pressure-gradient contribution to be included explicitly in addition to concentration-driven diffusion. The use of the convection-diffusion equation interface does not imply that bulk fluid convection occurs within the membrane; rather, the convection-like contribution represents the effective drift associated with the mechanical-pressure gradient in the chemical potential, as defined in Section 2.5.
The mechanical response was implemented using the Solid Mechanics interface. The Parallel Network Model and hydration-dependent material properties described in Section 2.3 were incorporated into the mechanical formulation, while the local transient water concentration was used to determine hygroscopic swelling according to Section 2.4. In the fully coupled model, the transport and mechanical fields were solved together, allowing the evolving concentration field to determine swelling and mechanical deformation while the resulting mechanical-pressure field contributed to the modified transport equation.

#### 2.6.1 Geometry, Initial Condition, and Transport Boundary Conditions

The computational geometry and boundary conditions are shown in Fig. 3. The membrane was initially assumed to be in equilibrium with the dry-side environment. Accordingly, the initial water concentration throughout the membrane domain was prescribed as

$$c_w(\mathbf{X}.0) = c_{w.dry} \tag{25}$$

where $c_{w.dry}$ is the equilibrium water concentration corresponding to the imposed dry-side relative humidity, determined from the water sorption relationship described in Section 2.2. At the beginning of the transient simulation, a water-activity gradient was imposed across the membrane thickness. The upper membrane surface was exposed to the wet-side condition, whereas the lower surface was maintained at the dry-side condition. The corresponding concentration boundary conditions were

$$c_w = c_{w.\ wet} \quad \text{on } \Gamma_{wet} \tag{26}$$

$$c_w = c_{w.\ dry} \quad \text{on } \Gamma_{dry} \tag{27}$$

where $c_{w.\ wet}$ and $c_{w.\ dry}$ are the equilibrium membrane water concentrations corresponding to the wet- and dry-side relative humidities, respectively. These concentrations were obtained from the membrane water sorption relationship described in Section 2.2. Thus, the externally imposed difference in water activity establishes a difference in equilibrium water concentration between the two membrane surfaces, providing the driving force for water transport across the membrane.

For the conditions considered in this study, the wet-side relative humidity ranged from 20% to 80%, whereas the dry-side relative humidity ranged from 0.5% to 20%, depending on the condition being simulated.

The two lateral boundaries were treated as impermeable to water transport:

$$\mathbf{n} \cdot \mathbf{j}_{\mathrm{w}} = 0 \quad \text{on } \Gamma_{lat} \tag{28}$$

where $\mathbf{n}$ is the outward unit normal vector and $\mathbf{j}_{\mathrm{w}}$ is the water flux. The no-flux condition prevents water transport through the lateral boundaries, such that net water transport occurs across the membrane thickness between the wet and dry surfaces.

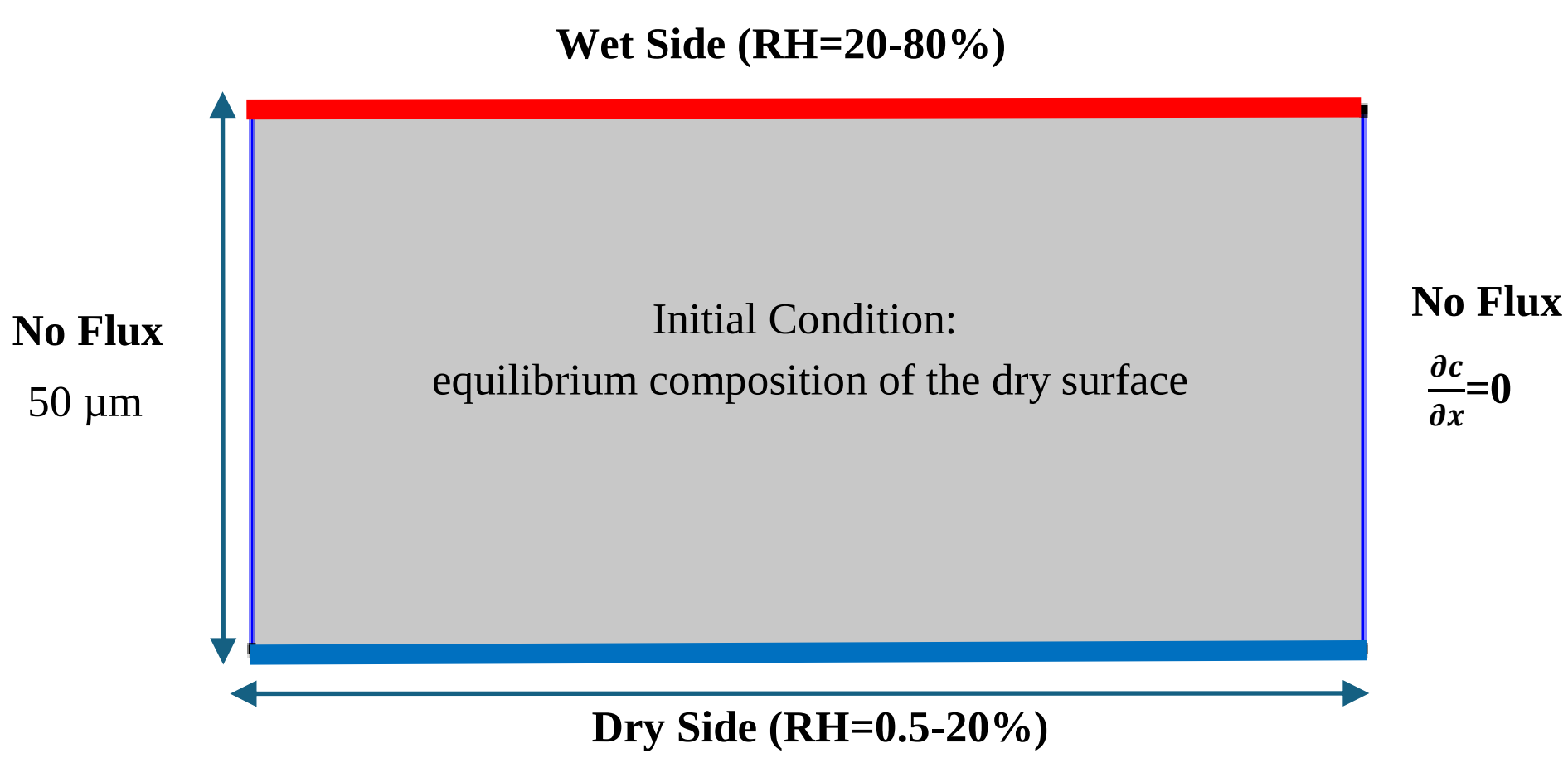

**Figure 3.** Two-dimensional computational geometry and boundary conditions used for the FAA-3-50 membrane simulations. The membrane has an initial thickness of 50 μm and an in-plane length of 100 μm. The membrane is initially equilibrated with the dry-side condition. At t>0, the upper surface is exposed to the wet-side water activity and the lower surface is maintained at the dry-side water activity. The lateral surfaces are impermeable to water transport. For the mechanical problem, the lateral boundaries are fixed, whereas the upper and lower surfaces are mechanically free.

### 2.6.2 Mechanical Boundary Conditions

The mechanical response of the membrane was solved using the Solid Mechanics interface. Both lateral boundaries of the membrane were fixed,

$$\mathbf{u} = \mathbf{0} \quad on\ \Gamma_{left} \cup \Gamma_{right} \tag{29}$$

where **u** is the displacement vector. The wet-side and dry-side surfaces were mechanically free and were therefore assigned traction-free boundary conditions,

$$\boldsymbol{\sigma}\mathbf{n} = \mathbf{0} \quad on\ \Gamma_{wet} \cup \Gamma_{dry} \tag{30}$$

where **σ** is the Cauchy stress tensor and **n** is the outward unit normal vector.

These boundary conditions constrain displacement at the lateral edges while permitting the upper and lower membrane surfaces to deform freely in response to hygroscopic swelling. The constraints consequently influence the spatial distribution of swelling-induced stresses, particularly near the fixed lateral boundaries and corner regions. The same mechanical boundary conditions were used in the one-way and fully coupled formulations.

### 2.6.3 Implementation of the Three Modeling Formulations

Three numerical formulations were evaluated using the same computational geometry, boundary conditions, and concentration-dependent water diffusivity.

In the uncoupled transport model, water transport was calculated using the concentration-dependent diffusivity $D(c_w)$ described in Section 2.2, while mechanical deformation and its influence on transport were excluded. Thus, although this formulation is mechanically uncoupled, the water diffusivity continues to vary locally with the evolving water concentration.

In the one-way coupled model, the transient water concentration field was supplied to the Solid Mechanics model. At each location within the membrane, the local value of $c_w$ was used to evaluate the hydration-dependent mechanical properties and the hygroscopic strain described in Sections 2.3 and 2.4. The resulting displacement and stress fields did not feed back into the transport calculation. Therefore, the coupling was unidirectional from water transport to mechanics.

Importantly, the hygroscopic response was evaluated locally throughout the membrane domain using the transient concentration field, rather than being prescribed only at the wet-side boundary.

As water penetrated through the membrane, the local swelling strain and associated mechanical response therefore evolved continuously in both space and time.

For the fully coupled model, the modified transport equation developed in Section 2.5 was implemented using COMSOL's Stabilized Convection-Diffusion Equation (scdeq) interface.

The scdeq and Solid Mechanics formulations were coupled through the evolving concentration and mechanical fields. The local water concentration determined hygroscopic swelling and the hydration-dependent mechanical response, while the resulting membrane deformation contributes to the transport equation through Eq. (20). This implementation provided the two-way feedback required by the fully coupled formulation described in Section 2.5.

Hygroscopic swelling resulted in a continuously evolving membrane configuration during the transient simulation. The local volumetric deformation was characterized by the determinant of the deformation gradient,

$$J = \det(\mathbf{F}) = \frac{V}{V_0} \tag{31}$$

where $J$ is the local volume ratio, $\mathbf{F}$ is the total deformation gradient defined in Section 2.5, $V$ is the current local volume, and $V_0$ is the corresponding reference volume. Accordingly, $J = 1$ represents the undeformed reference state, whereas $J > 1$ represents local volumetric expansion.

The volume ratio was evaluated throughout the membrane domain to characterize the spatial and temporal evolution of swelling-induced volumetric deformation. Because geometric nonlinearity was included in the Solid Mechanics formulation, the calculated displacement field defined the evolving spatial configuration of the membrane. Consequently, the transport and mechanical fields were evaluated within the coupled, deformation-dependent configuration, allowing changes in membrane geometry associated with swelling to be represented in the transient solution.

The local volume change was additionally incorporated explicitly into the prescribed transport boundary concentrations. The concentration imposed at the wet and dry membrane surfaces was updated according to

$$c_{w.\ BC}^{updated} = \frac{c_{w.\ BC}}{J_{BC}} \tag{32}$$

Where $c_{w.\ BC}$ is the equilibrium boundary concentration defined for the reference membrane state and $J_{BC}$ is the local volume ratio evaluated at the corresponding boundary. Thus, as the boundary region expanded during hydration ($J_{BC} > 1$), the prescribed water concentration was adjusted to account for the increased local volume.

Within the membrane bulk, the transient water concentration was determined by solving the stabilized convection-diffusion equation on the evolving membrane configuration. The displacement and volumetric deformation generated by hygroscopic swelling evolved throughout the domain and were coupled to the transport solution through the deformation-dependent geometry. In the fully coupled formulation, this geometrical feedback acted together with the mechanical-pressure contribution to the transport equation, allowing both deformation and the evolving mechanical state to influence water transport.

### 2.6.4 Mesh and Time-Dependent Solution

The computational domain was discretized using a structured finite-element mesh. Mesh refinement was concentrated across the membrane thickness to adequately resolve the water concentration gradients and the corresponding spatial variations in hygroscopic swelling and mechanical response. The same mesh was used for the uncoupled, one-way coupled, and fully coupled formulations to maintain consistency among the model comparisons.

All simulations were performed using a time-dependent study. The membrane was initially equilibrated at the dry-side condition, after which the wet-side boundary condition was imposed at $t=0$. The transient concentration and, for the coupled models, displacement fields were then solved simultaneously over time. The simulations were continued until the concentration distribution and water flux approached steady-state conditions.

For the fully coupled formulation, geometric nonlinearity was enabled to account for the finite deformation associated with hygroscopic swelling. At each time step, the water concentration field determined the local hygroscopic strain and hydration-dependent mechanical properties, while the resulting deformation and mechanical-pressure fields were returned to the transport formulation. This iterative coupling allowed the concentration and displacement fields to evolve consistently throughout the transient simulation.

### 2.6.5 Numerical Modeling Workflow

The overall numerical procedure is summarized in Fig. 4. All three formulations were initialized using the same membrane geometry, dry-side equilibrium concentration, concentration-dependent diffusivity, and imposed water-activity boundary conditions. The uncoupled model resolved water transport without mechanical deformation. The one-way coupled model subsequently incorporated the mechanical response to the evolving local water concentration without feedback to transport. Finally, the fully coupled model simultaneously solved the stabilized convection-diffusion and Solid Mechanics formulations, allowing the evolving concentration and mechanical fields to interact throughout the simulation.

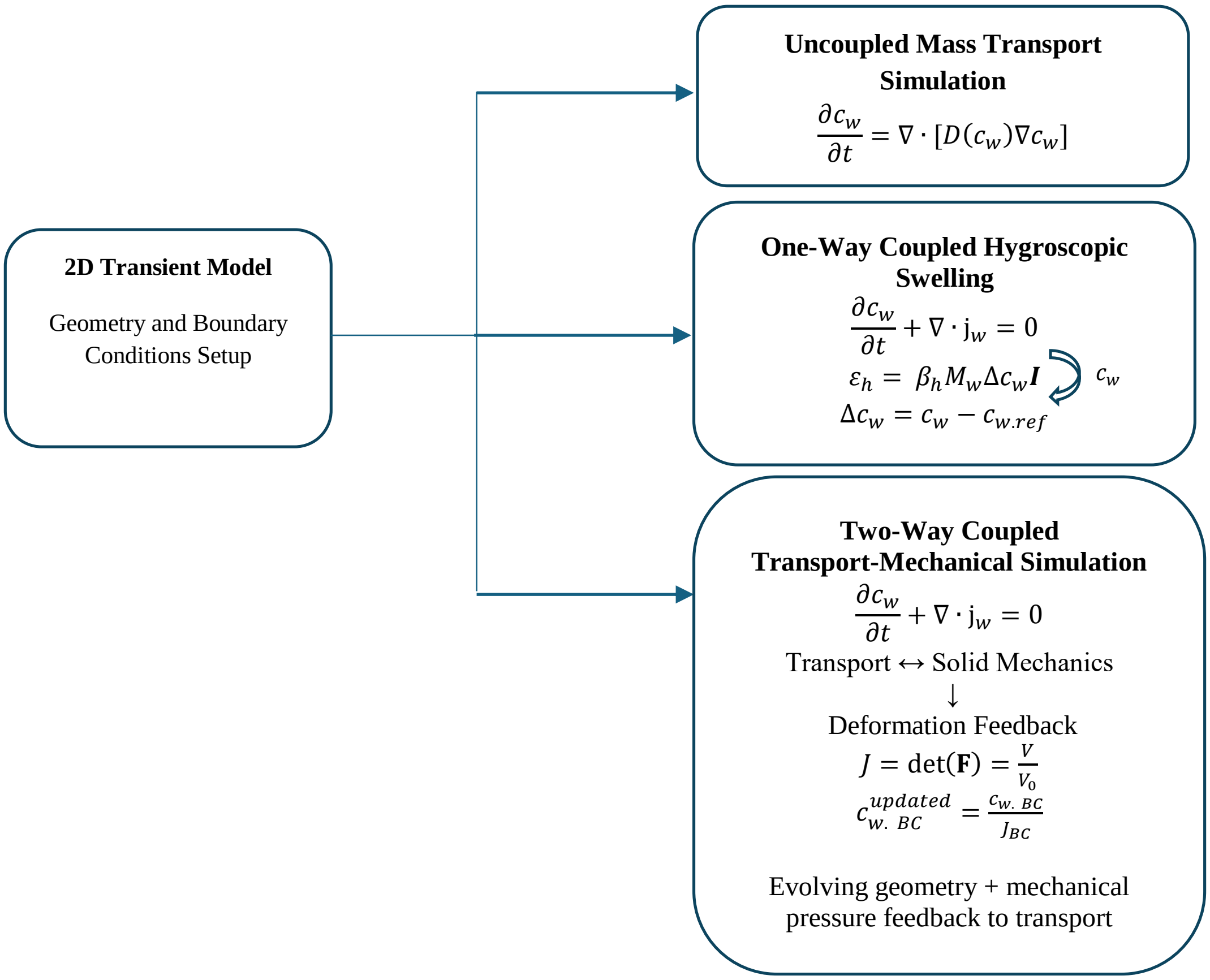

**Figure 4.** Numerical workflow for the uncoupled, one-way coupled, and fully coupled transport–mechanical formulations. The uncoupled model considers concentration-dependent water diffusion, the one-way model additionally calculates hygroscopic swelling and mechanical deformation from the local water concentration, and the fully coupled model allows deformation, mechanical pressure, and the evolving membrane configuration to feed back into water transport.

## 2.7 Materials

Fumasep FAA-3-50 anion exchange membranes in bromide form were obtained from Fuel Cell Store (Bryan, TX, USA). Potassium bicarbonate ($KHCO_3$, 99.7%) was purchased from Sigma-Aldrich (USA) and used as received. Deionized (DI) water was produced using a Millipore RiOs and A10 purification system.

### 2.7.1 Ion Exchange Procedure

Fumasep FAA-3-50 membranes were converted to the bicarbonate form by immersion in a 0.5 M $KHCO_3$ aqueous solution for 24 h at room temperature (20 mg membrane in 15 mL solution). The membranes were then transferred to a fresh $KHCO_3$ solution and equilibrated for an additional 24 h to promote further ion exchange. Afterward, the membranes were thoroughly rinsed with DI water to remove residual salts.

### 2.7.2 Water Vapor Permeability Measurements

Membranes were mounted onto brass support disks using Devcon epoxy and allowed to cure for a minimum of 12 h to ensure a leak-free seal. Membrane thickness was measured using a digital caliper (Mitutoyo, ±1 µm resolution).

Water vapor permeability was measured at 30 °C using a mixed-gas permeation system described in previous studies and shown in Fig. 5 [17]. The membrane was exposed to controlled humidity gradients by adjusting the relative humidity (RH) of the feed stream, while a dry sweep gas was used on the permeate side to maintain a driving force for transport. Water flux was determined from the steady-state vapor concentration in the permeate stream.

To ensure that the measured permeability reflected intrinsic membrane transport properties rather than external mass transfer limitations, multiple feed and sweep flow rates were tested. Conditions were selected such that concentration polarization effects were negligible, following established procedures [18].

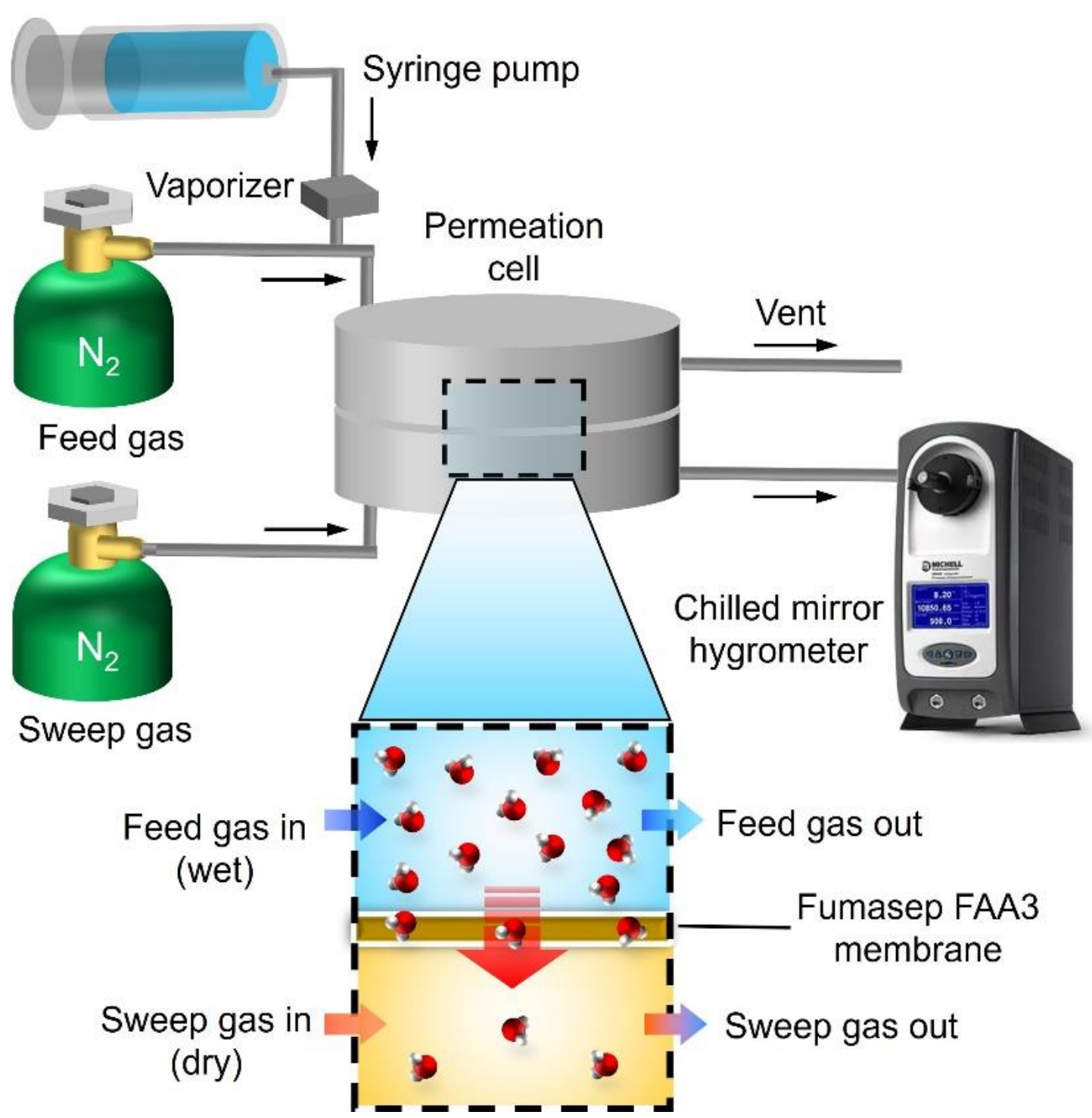

Figure 5. Schematic of the experimental setup used to measure water vapor transport through a Fumasep FAA-3 membrane.

## 3. Results and Discussion

### 3.1 Transient Evolution of Water Concentration

The transient water-transport response was first examined to establish how the imposed water-activity gradient develops within the membrane before considering the associated mechanical response and steady-state behavior. Figure 6 shows the evolution of the water concentration profile across the FAA-3-50 membrane for a representative condition with wet- and dry-side water activities of aH2O=0.59 and aH2O=0.031, respectively. The membrane was initially equilibrated with the dry-side environment before the higher water activity was imposed at the wet surface.

At early times, the increase in water concentration is confined primarily to the region adjacent to the wet surface, producing a steep concentration gradient through the membrane. As time progresses, water penetrates toward the dry side and the region of elevated concentration extends progressively across the membrane thickness. The concentration profiles consequently evolve from a strongly localized initial response toward a persistent through-thickness gradient.

The transient response is nonlinear because the water diffusivity varies with the local water concentration. As described in Section 2.2, D(cw) increases with hydration. The initially dry portion of the membrane therefore exhibits a lower diffusivity than regions that have already absorbed water. As the membrane hydrates, the local increase in D(cw) changes the rate of further water redistribution, causing both the magnitude and shape of the concentration gradient to evolve with time.

The differences between successive concentration profiles become progressively smaller as the simulation proceeds, and the concentration field approaches a steady distribution by approximately 400 s for the representative condition shown. Importantly, the steady-state concentration does not become spatially uniform because different water activities remain imposed at the two membrane surfaces. Instead, a finite concentration gradient persists across the membrane at steady state.

The timescale associated with this response is controlled primarily by transport across the membrane thickness. A characteristic diffusive timescale scales as

$$t_D \sim \frac{L^2}{D} \tag{33}$$

where L is the membrane thickness and D is a characteristic water diffusivity. In the present system, however, D is concentration dependent, so a single diffusivity cannot describe the complete transient process. Water initially moves through a relatively low-diffusivity dry membrane, while progressive hydration increases D(cw) and alters the local transport rate. The characteristic transport timescale therefore evolves during hydration rather than remaining constant.

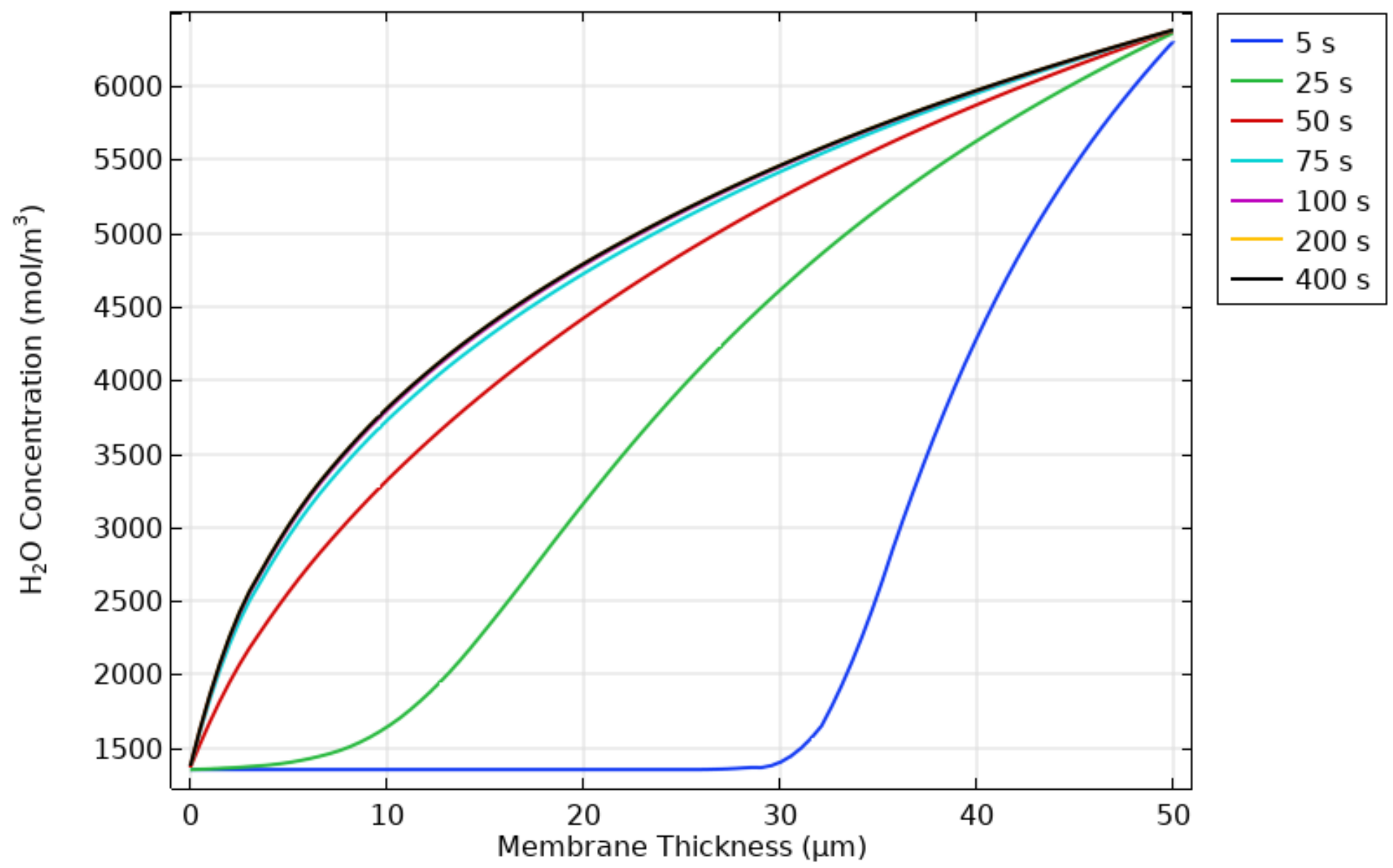

**Figure 6.** Temporal evolution of water concentration profiles across the FAA-3-50 membrane under a representative imposed water-activity gradient of aH2O, wet=0.59 and aH2O,dry=0.031. The membrane is initially equilibrated with the dry-side condition.

Figure 6 presents a representative condition used to illustrate the transient behavior of the coupled system. The experimental validation discussed later includes different feed and permeate humidity conditions; the measured permeate RH, for example, varies from approximately 0.5% to 19% across the experimental series.

### 3.2 Coupled Concentration and Deformation Response

The spatial and temporal relationship between hydration and membrane deformation is summarized in Fig. 7. Figure 7(a) shows the steady-state water concentration distribution corresponding to the transient response discussed above. The concentration remains highest near the wet surface and decreases toward the dry surface, reflecting the water-activity difference maintained across the membrane.

The steady-state concentration distribution is nonlinear rather than linear. This behavior is a direct consequence of the concentration-dependent diffusivity: because the local diffusivity changes with hydration, an approximately constant steady-state water flux does not require a constant concentration gradient. Instead, the local concentration gradient adjusts according to the spatial variation in D(cw), giving rise to the curved steady-state concentration distribution.

The associated steady-state displacement field is shown in Fig. 7(b). Hygroscopic deformation closely follows the spatial variation in water concentration because the local swelling strain is calculated from the local change in water concentration throughout the membrane. Regions near the wet surface experience greater water uptake and therefore greater swelling, whereas the less hydrated region near the dry surface undergoes a smaller mechanical response. The resulting displacement field is consequently nonuniform across the membrane.

The temporal displacement histories at the wet and dry surfaces are shown in Figs. 7(c) and 7(d), respectively. At the wet surface, displacement begins to increase rapidly after application of the higher water activity because the local concentration responds immediately to the changed boundary condition. The displacement subsequently approaches a plateau as the local concentration and swelling response approach steady values. The maximum displacement is approximately 1.5 μm for the representative condition.

In contrast, the dry-side displacement develops more gradually. The dry surface is not exposed directly to the abrupt increase in water activity; its mechanical response develops as the concentration field progresses through the membrane interior. The difference between Figs. 7(c) and 7(d) therefore reflects the finite time required for water redistribution across the membrane and confirms that the mechanical response evolves locally with the transient hydration field.

The concentration and displacement fields both approach steady behavior on a timescale of several hundred seconds for the representative case. Although the viscoelastic constitutive model introduces an intrinsic rate dependence to the mechanical response, the spatial delay between the wet- and dry-side displacement histories demonstrates that the overall coupled timescale is strongly linked to water transport through the membrane. Thus, the transient response reflects both hydration-dependent diffusion and the time-dependent mechanical behavior of the polymer network.

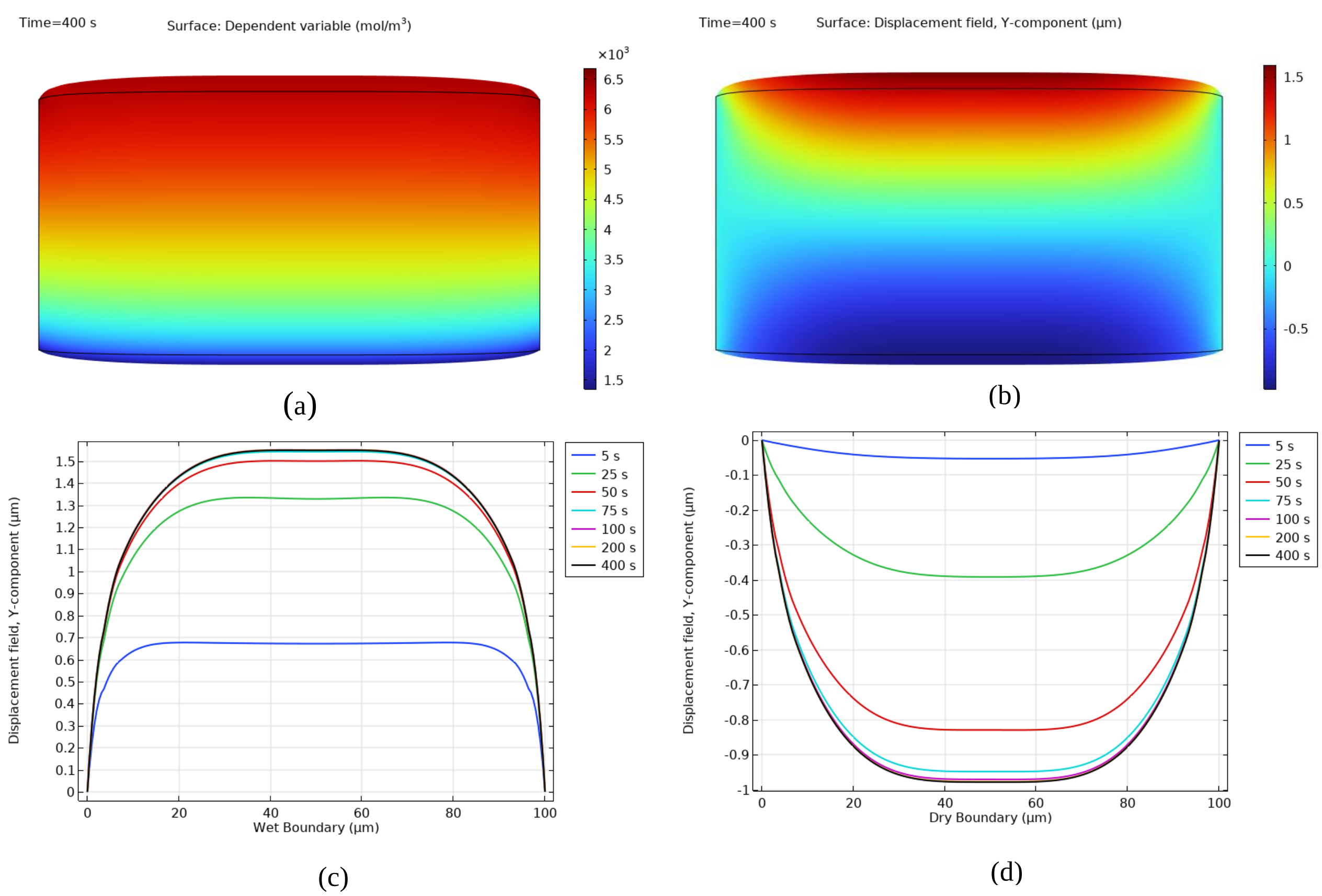

**Figure 7.** Coupled water concentration and deformation response of the FAA-3-50 membrane under the representative imposed water-activity gradient: **(a)** steady-state water concentration distribution at t=400 s; **(b)** steady-state displacement distribution at t=400 s; **(c)** temporal evolution of displacement at the wet surface; and **(d)** temporal evolution of displacement at the dry surface.

### 3.3 Swelling-Induced Mechanical Pressure

The nonuniform swelling generated by the concentration gradient also produces an internal mechanical-pressure field. Figure 8 shows the pressure distribution at t=400 s, after the concentration and deformation fields have approached steady behavior. The pressure field is strongly nonuniform near the lateral boundaries and corners, whereas the membrane interior exhibits a comparatively smoother spatial distribution.

The location of the largest pressure magnitudes is closely related to the mechanical boundary conditions. The two lateral boundaries are fixed, while the wet and dry surfaces are mechanically free. The fixed boundaries prevent the polymer from undergoing its preferred in-plane hygroscopic expansion, producing locally elevated stresses and mechanical pressure. This constraint is particularly pronounced in the corner regions, where the fixed lateral boundaries intersect the freely deforming upper and lower surfaces. Away from these constrained regions, swelling can be accommodated more readily, resulting in a smoother mechanical-pressure field through most of the membrane interior.

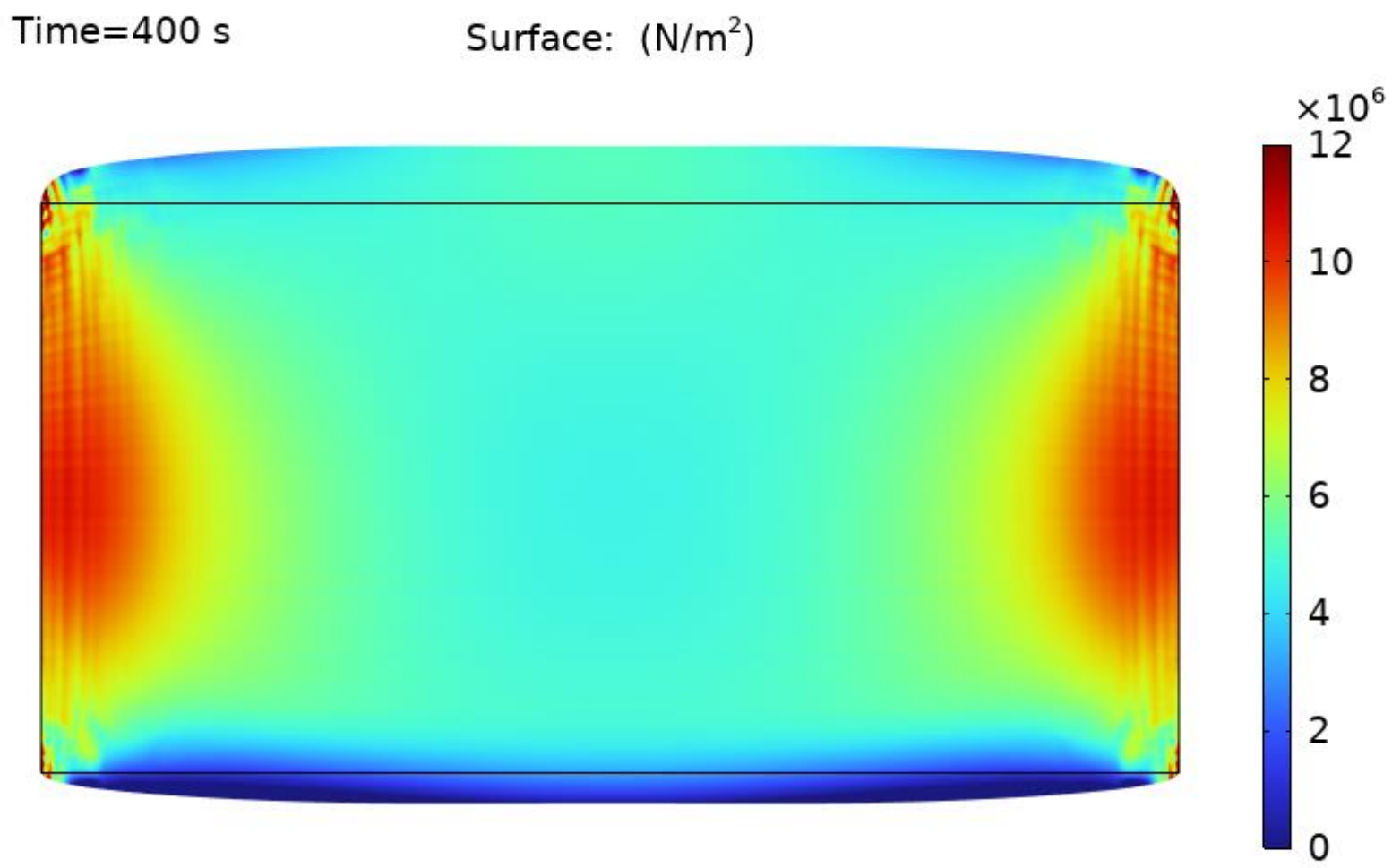

**Figure 8.** Spatial distribution of the swelling-induced mechanical pressure within the FAA-3-50 membrane at t=400 s. Elevated pressure magnitudes near the lateral boundaries and corner regions result from the fixed mechanical constraints that locally restrict hygroscopic expansion.

Although the local pressure reaches values on the order of MPa near constrained regions, its contribution to water transport is governed by the spatial pressure gradient rather than by the absolute pressure magnitude. The comparatively gradual pressure variation across most of the membrane therefore limits the direct mechanically induced contribution to water flux. This interpretation is consistent with the Peclet-type analysis introduced in Section 2.5, for which Pem<1 under the simulated conditions. The transport therefore remains dominated by the concentration-gradient contribution even in the presence of substantial local swelling-induced stresses.

The small magnitude of the direct pressure-gradient contribution does not imply that mechanical coupling is unimportant. Instead, the model distinguishes between direct and indirect mechanical effects. Hygroscopic swelling changes the membrane configuration and local volume and modifies the concentration boundary conditions through the volume-ratio correction described in Section 2.6. The average volume ratio at the wet boundary reaches approximately 1.15 for the representative condition, indicating appreciable local volumetric expansion.

Thus, the primary influence of mechanical coupling under the present conditions arises through deformation-induced modification of the membrane configuration and transport environment rather than through a dominant mechanical-pressure-driven water flux.

### 3.4 Model Validation

After establishing the transient concentration, deformation, and mechanical-pressure fields, the predictive capability of the model was evaluated against experimental water-vapor transport measurements. Validation was considered in three stages. First, predictions from the uncoupled concentration-dependent transport model were compared with those from the fully coupled formulation to isolate the influence of mechanical feedback on water transport. Second, both the fully coupled and uncoupled predictions were quantitatively compared with the measured steady-state permeability and water flux over the investigated humidity range. Finally, the model–experiment comparison was evaluated in the context of experimental and model uncertainty.

#### 3.4.1 Effect of Mechanical Coupling on Predicted Water Flux

The effect of mechanical feedback on water transport was isolated by comparing the uncoupled transport model with the fully coupled transport–mechanical formulation. Importantly, both models employ the same concentration-dependent water diffusivity, D(cw). The uncoupled model therefore already accounts for the hydration dependence of intrinsic water mobility, and differences between the two predictions can be attributed to the additional deformation-induced feedback included in the fully coupled formulation.

In the uncoupled model, water transport is governed by the species conservation equation, Eq. (1), together with the concentration-dependent Fickian flux given by Eq. (5),

$$\mathbf{j}_w = -D(c_w)\nabla c_w$$

In the fully coupled formulation, transport is governed by Eq. (20), which retains the same D(cw) relationship while additionally incorporating the mechanical-pressure contribution to the chemical potential and the effects of the evolving membrane configuration. Thus, the comparison does not contrast constant and concentration-dependent diffusivities; rather, it isolates the additional influence of chemo-mechanical coupling.

Figure 9 compares the experimental steady-state water flux with predictions from the uncoupled concentration-dependent transport model and the fully coupled model. The uncoupled D(cw) formulation captures the overall increase in water flux with increasing feed humidity, demonstrating that the hydration dependence of diffusivity accounts for a substantial portion of the observed nonlinear transport behavior. However, the uncoupled model increasingly overpredicts the measured water flux as the feed humidity increases, with the discrepancy becoming more pronounced at the higher-humidity conditions where membrane hydration and swelling are greater.

The fully coupled formulation predicts lower water fluxes than the uncoupled D(cw) model at these higher-hydration conditions and provides closer agreement with the experimental measurements over most of the investigated humidity range. Because the Peclet-type analysis in Section 3.3 indicates that the explicit mechanical-pressure contribution to water flux is small, the difference between the uncoupled and fully coupled predictions is attributed primarily to the indirect effects of swelling-induced deformation, including changes in membrane geometry, local volume, and boundary concentrations.

These results demonstrate that concentration-dependent diffusivity alone is not sufficient to quantitatively reproduce water transport over the full range of hydration conditions considered. At lower hydration, where swelling is comparatively limited, the uncoupled formulation provides a reasonable approximation of the measured transport behavior. As hydration and deformation increase, however, neglecting mechanical feedback leads to a progressively greater overprediction of water flux. The uncoupled formulation therefore becomes less representative of the membrane response as swelling becomes significant, indicating that water transport and mechanical deformation cannot be treated as independent processes under these conditions.

The comparison also distinguishes two sources of nonlinearity in the membrane response. The concentration dependence of D(cw) governs much of the intrinsic hydration dependence of water mobility, whereas chemo-mechanical coupling accounts for the additional modification of transport associated with swelling and deformation. Thus, while D(cw) is essential for describing hydration-dependent diffusion, incorporation of mechanical coupling is necessary for quantitatively reproducing water transport in FAA-3-50 under conditions where hydration-induced deformation is appreciable.

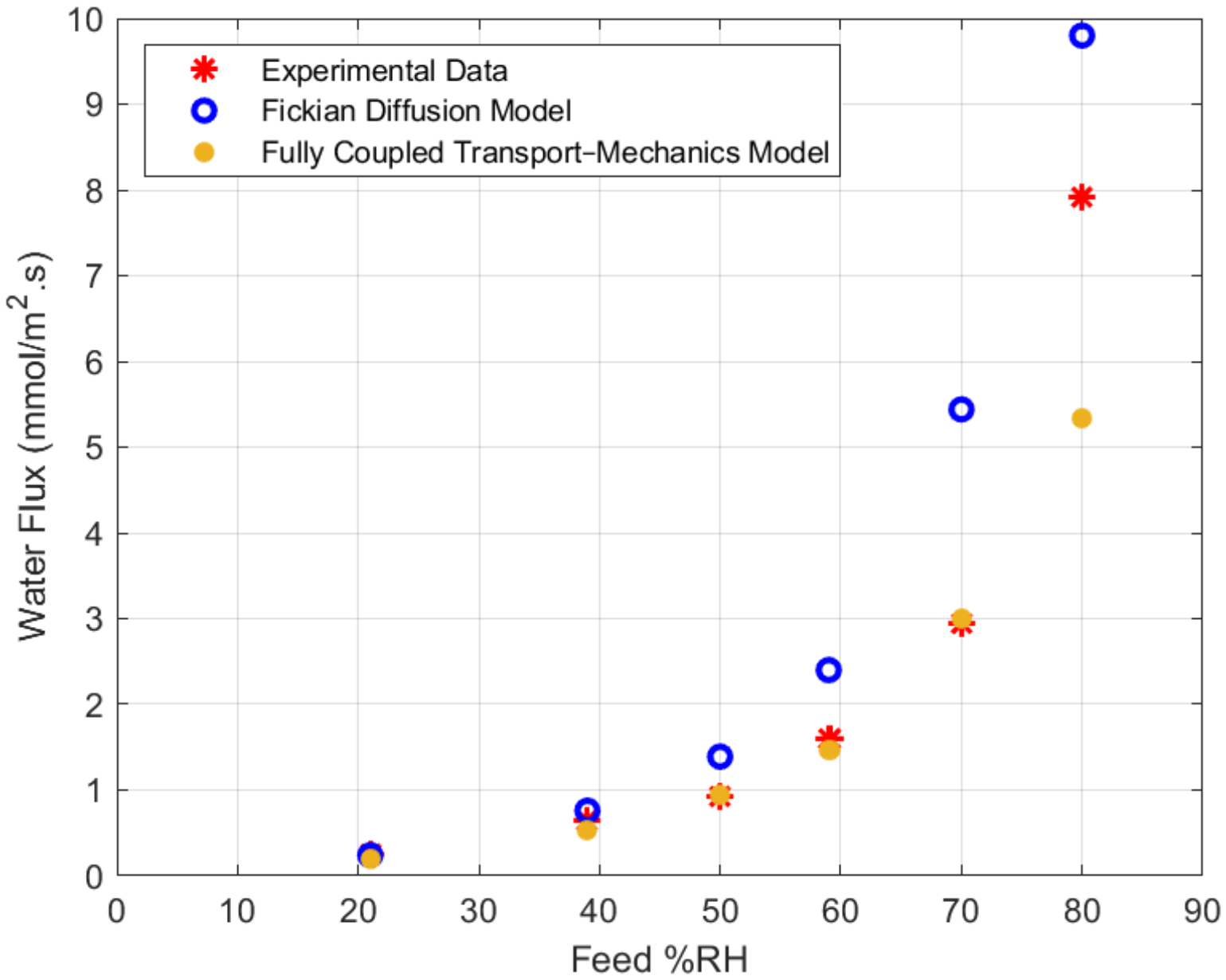

**Figure 9.** Experimental steady-state water flux as a function of feed relative humidity compared with predictions from the uncoupled concentration-dependent diffusion model [Eqs. (1) and (5)] and the fully coupled transport–mechanical model [Eq. (20)]. The same concentration-dependent diffusivity, D(cw), is used in both numerical formulations.

### 3.4.2 Quantitative Comparison with Experimental Measurements

The predictive performance of the fully coupled model was further evaluated by comparing the simulated steady-state water permeability and flux with the corresponding experimental measurements over feed relative humidities ranging from approximately 20% to 80%. The experimental and simulated results are summarized in Table 2.

The fully coupled model shows the closest agreement with experiment at 50% RH, with a relative deviation of 2.2%. Agreement remains strong at 59% and 70% RH, where the deviations are 8.2% and 8.5%, respectively. Larger deviations of 26% and 17% are observed at 21% and 39% RH; however, the measured fluxes at these conditions are relatively low and have larger fractional experimental uncertainties. These deviations should therefore be interpreted in the context of the associated measurement variability. The largest discrepancy occurs at 80% feed RH, where the model underpredicts the measured water flux by approximately 33%.

The discrepancy at 80% RH should not be attributed to a single mechanism based on the present results. At high hydration, the predicted transport response depends on the concentration-dependent diffusivity, water sorption behavior, swelling response, and mechanical properties, while experimental uncertainty may also contribute. Additional characterization of these hydration-dependent properties at high water activity would help determine whether the remaining discrepancy arises primarily from parameter uncertainty or indicates a limitation of the current constitutive description.

Overall, the fully coupled model captures the measured humidity dependence of water transport, with particularly close quantitative agreement between 50% and 70% RH. The larger discrepancy at 80% RH identifies the highly hydrated regime as the primary condition for further model refinement and experimental characterization. Together with the comparison in Fig. 9, these results support the inclusion of mechanical coupling for quantitatively predicting water transport in FAA-3-50 as hydration and swelling increase.

**Table 2. Experimental measurements and fully coupled model predictions of steady-state water permeability and water flux for FAA-3-50 under the investigated humidity conditions.**

| Feed RH (%) | Permeate RH (%) | Experimental permeability (Barrer) | Simulated permeability (Barrer) | Experimental flux (mmol m⁻² s⁻¹) | Simulated flux (mmol m⁻² s⁻¹) | Relative deviation (%) |
|---|---|---|---|---|---|---|
| **21 ± 1** | 0.5 ± 0.2 | $(4.4 \pm 1.3) \times 10^4$ | $3.3 \times 10^4$ | 0.26 ± 0.08 | 0.19 | 26 |
| **39 ± 1.5** | 1.3 ± 0.6 | $(6.1 \pm 3.0) \times 10^4$ | $5.0 \times 10^4$ | 0.65 ± 0.30 | 0.54 | 17 |
| **50 ± 1.5** | 3.5 ± 1.5 | $(9.4 \pm 0.26) \times 10^4$ | $9.5 \times 10^4$ | 0.92 ± 0.08 | 0.94 | 2.2 |
| **59 ± 1.5** | 3.1 ± 0.9 | $(1.0 \pm 0.32) \times 10^5$ | $9.2 \times 10^4$ | 1.59 ± 0.50 | 1.5 | 8.2 |
| **70 ± 1.5** | 6.5 ± 0.3 | $(1.6 \pm 0.06) \times 10^5$ | $1.5 \times 10^5$ | 2.95 ± 0.10 | 2.7 | 8.5 |
| **80 ± 1** | 19 ± 1.8 | $(4.6 \pm 0.19) \times 10^5$ | $3.1 \times 10^5$ | 7.93 ± 0.30 | 5.3 | 33 |

**Note:** Relative deviation was calculated as

Relative deviation (%)=$|J_{sim} - J_{exp}| \times 100 / J_{exp}$

### 3.4.3 Uncertainty Analysis

The agreement between the fully coupled model and experiment was further evaluated by considering uncertainty in the experimental measurements and model predictions. Figure 10 presents the experimental steady-state water fluxes together with the fully coupled predictions and their corresponding uncertainty ranges.

The experimental error bars account for variability in the measured transport response. The RH and permeability uncertainties are reported as the standard deviation of triplicate measurements. The model predictions agree closely with the experimental measurements through much of the intermediate humidity range, consistent with the quantitative comparison in Table 2.

At the highest feed humidity, the central model prediction lies below the measured value, although the reported model prediction interval overlaps the experimental result. This overlap indicates that the difference between the central values should be considered in the context of the uncertainty associated with both the experiment and model. It does not, however, establish that uncertainty alone accounts for the discrepancy at 80% RH.

The absolute width of the model prediction interval increases as the magnitude of water flux increases. This trend should not be interpreted directly as an increase in relative uncertainty because the predicted flux itself also increases substantially with feed humidity. A larger uncertainty interval in absolute units may therefore correspond to a similar fractional uncertainty. A normalized uncertainty analysis would be required to determine whether the relative predictive uncertainty systematically increases with hydration.

The 80% RH condition nevertheless represents the most stringent test of the present model because it exhibits the largest difference between the central prediction and the experimental measurement. Further characterization of water diffusivity, sorption behavior, swelling, and mechanical properties under highly hydrated conditions would help constrain the model parameters in this regime and determine whether refinement of the coupled constitutive formulation is required.

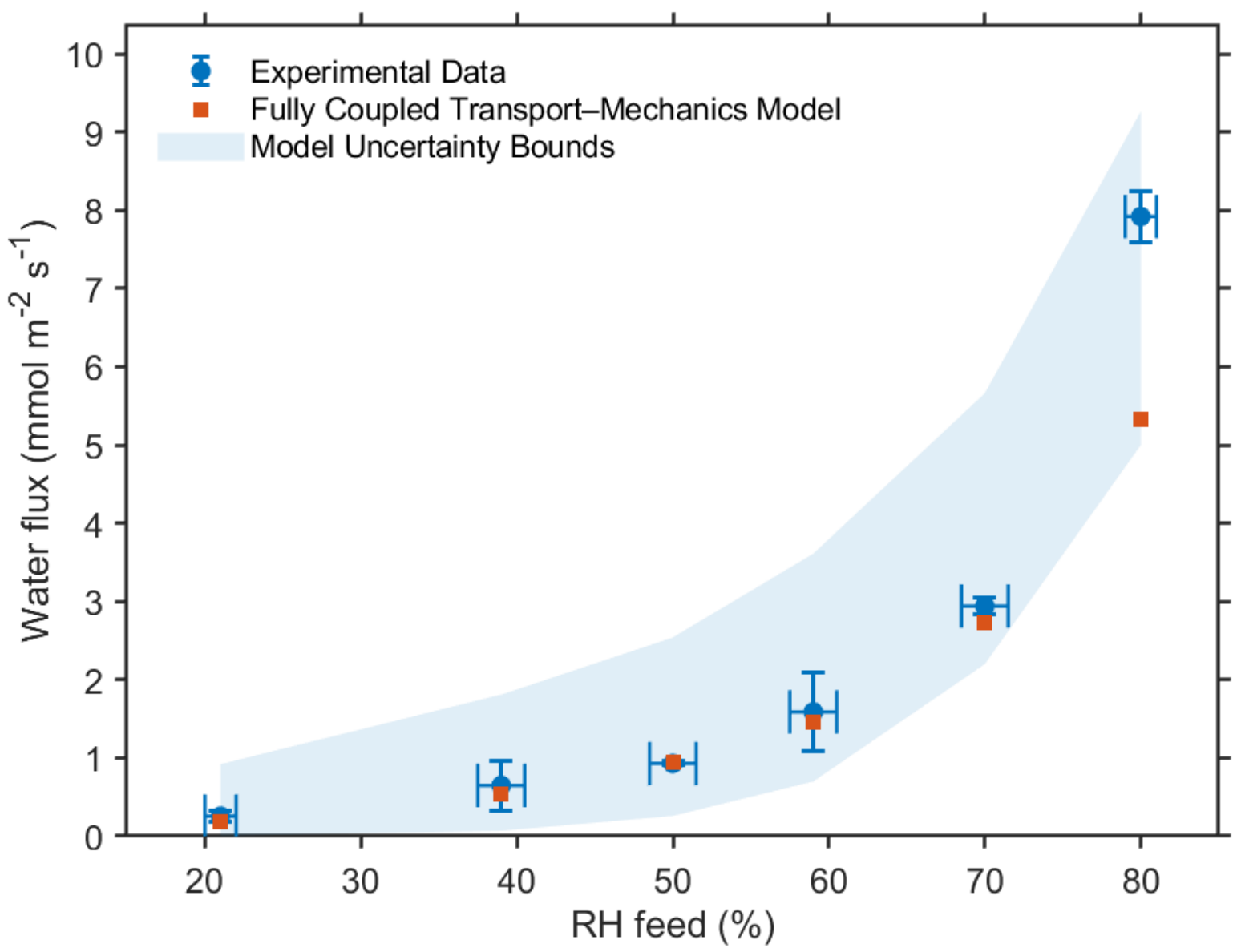

Figure 10. Comparison of experimental steady-state water flux and fully coupled model predictions as a function of feed relative humidity, including experimental measurement uncertainty and the model prediction interval. Experimental error bars represent the standard deviation of triplicate measurements, and the shaded region indicates the model prediction interval.

## 4. Conclusions

This study developed a transient, fully coupled transport–mechanical framework to investigate water transport and deformation in Fumasep FAA-3-50 anion exchange membranes subjected to water-activity gradients. By incorporating concentration-dependent water diffusivity, hygroscopic swelling, and the nonlinear mechanical response of the membrane, the model captures the bidirectional interaction between hydration and deformation.

Comparison of the uncoupled and fully coupled formulations using the same concentration-dependent diffusivity demonstrated that hydration-dependent diffusivity accounts for much of the nonlinear water-transport behavior, but does not fully describe the membrane response as hydration and swelling increase. The fully coupled simulations showed that mechanical feedback modifies water transport primarily through swelling-induced changes in membrane geometry, local volume, and concentration distribution. In contrast, the direct contribution of mechanical-pressure gradients to water flux remained small under the conditions investigated, indicating that transport was predominantly diffusion-driven. These results distinguish the intrinsic effect of hydration on water mobility from the additional influence of deformation on membrane transport.

The fully coupled model reproduced the experimentally observed dependence of steady-state water transport on hydration, with particularly close agreement at intermediate hydration conditions. The remaining discrepancy at the highest feed humidity indicates that the highly hydrated regime requires further characterization, particularly of water diffusivity, sorption behavior, swelling, and mechanical properties.

Overall, this work demonstrates that water transport and mechanical deformation in hydrated anion exchange membranes become increasingly interdependent as swelling becomes appreciable. The proposed framework provides a mechanistic basis for identifying when concentration-dependent diffusion is sufficient and when transport-mechanical feedback must be considered. Although the present study is limited to water transport, resolving this hydration–mechanics interaction provides a foundation for future modeling of moisture-facilitated membrane processes in which water-activity gradients influence membrane transport and separation behavior.

## Appendix

Table A.1. List of symbols

| Symbol | Description / units |
|---|---|
| $\mathbf{T}$ | Absolute temperature (K) |
| $\mathbf{R}$ | Universal gas constant (J mol$^{-1}$ K$^{-1}$) |
| $\boldsymbol{\rho}$ | Membrane density (kg m$^{-3}$) |
| $\mathbf{E}$ | Young’s modulus (Pa) |
| $\boldsymbol{\nu}$ | Poisson’s ratio |
| $\mathbf{u}$ | Displacement vector (m) |
| $\boldsymbol{\sigma}$ | Cauchy stress tensor (Pa) |
| $\boldsymbol{\sigma}_A$, $\boldsymbol{\sigma}_B$ | Cauchy stress contributions of mechanical Networks A and B (Pa) |
| $\mathbf{f}_v$ | Body force per unit volume (N m$^{-3}$) |
| $\boldsymbol{\varepsilon}_{tot}$ | Total strain tensor |
| $\boldsymbol{\varepsilon}_{mech}$ | Mechanical strain tensor |
| $\boldsymbol{\varepsilon}_h$ | Hygroscopic strain tensor |
| $\mathbf{I}$ | Second-order identity tensor |
| $\mathbf{F}$ | Total deformation gradient |
| $\mathbf{F}_m$ | Mechanical deformation gradient |
| $\mathbf{F}_s$ | Swelling deformation gradient |
| $\mathbf{T}_R$ | First Piola–Kirchhoff stress tensor (Pa) |
| $\mathbf{G}_1$, $\mathbf{G}_2$ | Hydration-dependent shear moduli of Networks A and B (MPa) |
| $\hat{\boldsymbol{\tau}}$ | Effective viscous flow resistance of Network B (MPa) |
| $\boldsymbol{\xi}$ | Bergström–Boyce flow parameter |
| $\mathbf{c}_{BB}$ | Bergström–Boyce stress-dependence parameter |
| $\mathbf{m}_{BB}$ | Bergström–Boyce flow exponent |
| $\mathbf{c}_w$ | Local water concentration (mol m$^{-3}$) |
| $\mathbf{c}_{w,ref}$ | Reference water concentration for the strain-free state (mol m$^{-3}$) |
| $\mathbf{c}_{w,wet}$; $\mathbf{c}_{w,dry}$ | Equilibrium membrane water concentrations at wet and dry surfaces (mol m$^{-3}$) |
| $\boldsymbol{\Delta}\mathbf{c}_w$ | Change in local water concentration relative to $c_{w,ref}$ (mol m$^{-3}$) |
| $\mathbf{c}_{w,BC}$ | Equilibrium boundary water concentration in the reference state (mol m$^{-3}$) |
| $\mathbf{c}_{w,BC\ updated}$ | Deformation-adjusted prescribed boundary water concentration (mol m$^{-3}$) |
| $\mathbf{J}_{BC}$ | Local volume ratio at the boundary |
| $\mathbf{j}_w$ | Water molar flux vector (mol m$^{-2}$ s$^{-1}$) |
| $\mathbf{S}_w$ | Water source/sink term (mol m$^{-3}$ s$^{-1}$) |
| $\mathbf{D}(\mathbf{c}_w)$ | Concentration-dependent water diffusivity (m$^2$ s$^{-1}$) |
| $\boldsymbol{\mu}_w$ | Chemical potential of water (J mol$^{-1}$) |
| $\boldsymbol{\mu}_w^0$ | Reference chemical potential of water (J mol$^{-1}$) |
| $\mathbf{m}_w$ | Water mobility |

| Symbol | Description / units |
|---|---|
| $\mathbf{M_w}$ | Molar mass of water (kg mol$^{-1}$) |
| $\mathbf{\Omega_w}$ | Molar volume of water (m$^3$ mol$^{-1}$) |
| $\mathbf{p}$ | Internal mechanical pressure (Pa) |
| $\mathbf{v_m}$ | Effective mechanically induced drift velocity (m s$^{-1}$) |
| $\mathbf{j_{w,m}}$ | Mechanically induced contribution to water flux (mol m$^{-2}$ s$^{-1}$) |
| $\mathbf{Pe_m}$ | Peclet-type number for mechanical drift relative to diffusion (–) |
| $\mathbf{\beta_h}$ | Hygroscopic swelling coefficient (m$^3$ kg$^{-1}$) |
| $\mathbf{J}$ | Local volume ratio, det(F) = $V/V_0$ |
| $\mathbf{V; V_o}$ | Current and reference local volumes (m$^3$) |
| $\mathbf{L}$ | Initial membrane thickness (m) |
| $\mathbf{W}$ | In-plane computational length (m) |
| $\mathbf{n}$ | Outward unit normal vector |
| $\mathbf{\Gamma_{wet}; \Gamma_{dry}}$ | Wet- and dry-side membrane boundaries |
| $\mathbf{\Gamma_{lat}}$ | Impermeable lateral boundaries |
| $\mathbf{\Gamma_{left}; \Gamma_{right}}$ | Mechanically fixed lateral boundaries |
| $\mathbf{t}$ | Time (s) |
| $\mathbf{t_D}$ | Characteristic diffusion timescale (s) |
| $\mathbf{a_{H_2O}}$ | Water activity |
| $\mathbf{RH}$ | Relative humidity (%) |